\documentclass[runningheads,a4paper]{llncs}

\usepackage{amssymb}
\usepackage{wrapfig}
\usepackage{subfigure}
\usepackage{graphicx}
\graphicspath{ {images/} }

\usepackage{url}
\urldef{\mailsa}\path|{dcalacci, orenled, sandy}@media.mit.edu|
\urldef{\mailsb}\path|shrier@mit.edu|
\newcommand{\keywords}[1]{\par\addvspace\baselineskip
\noindent\keywordname\enspace\ignorespaces#1}

\usepackage{color}
\definecolor{Orange}{rgb}{1,0.5,0}

\begin{document}

\mainmatter  

\title{Breakout: An Open Measurement and Intervention Tool for Distributed Peer Learning Groups}

\titlerunning{Breakout: An Open Tool for Distributed Peer Learning Groups}

%
%
\author{Dan Calacci\inst{1}\and Oren Lederman\inst{1}\and David Shrier\inst{2}\and Alex (Sandy) Pentland\inst{1}}
\authorrunning{Breakout: An Open Tool for Distributed Peer Learning Groups}

\institute{MIT Media Lab, 75 Amherst Street, Cambridge, MA, 02139
\mailsa\\
\and
MIT Connection Science, 77 Massachusetts Avenue, Cambridge, MA, 02139\\
\mailsb\\
\url{}}

%
%

\toctitle{Lecture Notes in Computer Science}
\tocauthor{Authors' Instructions}
\maketitle

\begin{abstract}
We present Breakout, a group interaction platform for online courses that enables the creation and measurement of face-to-face peer learning groups in online settings. 
Breakout is designed to help students easily engage in synchronous, video breakout session based peer learning in settings that otherwise force students to rely on asynchronous text-based communication. 
The platform also offers data collection and intervention tools for studying the communication patterns inherent in online learning environments. 
The goals of the system are twofold: to enhance student engagement in online learning settings and to create a platform for research into the relationship between distributed group interaction patterns and learning outcomes.

\end{abstract}

\section{Introduction}
In recent years, the web has enabled an educational revolution in the form of massively open online courses, or MOOCs, courses that are interacted with solely on the web. 
While MOOCs have enabled classes to reach larger enrollment numbers than traditional classrooms can accomodate, and have increased the public reach of higher education, 
the nature of student engagement and patterns of student communication in online settings is still unclear.
Anecdotally, students and faculty have complained about the community features of extant MOOC platforms.
Online learning has transformed the paradigm of student-student and student-teacher interaction in the classroom, creating two distinct paradigms for online learning communication: synchronous and asychronous. \cite{ref1}. 
As online learning grows in popularity, studying how these communication paradigms impact student communication patterns, learning outcomes, and retention is increasingly important.

The primary means of peer communication in online learning environments is through asynchronous forums, a communication modality that has been called "an essential piece of online learning" \cite{Coetzee_2014}  \cite{mackness2010ideals}. 
In small-scale online classes, however, a combination of asynchronous and synchronous communication modalities in has been shown to maximize students' personal engagement \cite{ohlund2000impact}. 
By adding synchronous chat to an otherwise fully asynchronous communication system, students report experiencing "significantly more social presence than students using email", suggesting that synchronous communication systems in online learning platforms may have a positive net effect on student engagement \cite{Oztok_2013}. 
In some contexts, synchronous text communication also improves student learning outcomes in online environments \cite{Kuyath:2008:SPI:1559493}.

We expect synchronous video-based communications to extend and amplify the benefits of synchronous text-based chat. 
Synchronous breakout sessions have the potential to create a richer peer learning experience. In a 2002 study on the effect of different communication methods on cooperation, Bos \textit{et al.} found that text-based communication modalities are associated with the lowest levels of cooperation in comparison to audio, video, and face to face interaction. Video-based methods are closest to replicating face to face communication, the modality that was associated with the highest cooperation levels \cite{bos2002effects}.

Providing students with the option of synchronous video and audio based communication also creates an outlet for studying the group dynamics of online learning groups. 
Group dynamics refers to the study of behavioral and psychological processes that occur within a group \cite{forsyth2009group}.
There is a meaningful body of past work on computationally measuring the group dynamics of teams, but they have been generally limited to lab-based experiments on groups with prescribed team sizes and meeting goals \cite{kim2008meeting}.
In this paper, we seek to present a tool that will expand the potential subject pool of these experiments to distributed groups working in small teams over online video chat.

To explore the effect of synchronous video-based communication on online learning groups, and to provide a research tool to study real-life group dynamics data, we have created Breakout, an online interaction platform that enables students to engage in distributed group video chats. 
Breakout is able to collect data on group interactions by analyzing the communication patterns between team members, and offers a modular visualization system for researchers interested in studying the effect of real-time interventions on distributed team dynamics.

\section{Related Work}
The increased availability of inexpensive hardware and the ease of creating customized visualizations has increased the accessibility of measuring and influencing team communication. 
Past work on small team dynamics and meeting feedback is split between real-time feedback systems and post-hoc systems, which include meeting summarization systems. DiMicco et al. used a post-hoc visualization system to encourage speakers to reflect on their social interaction \cite{dimicco2006using}. 
Later iterations on the same system provided a real-time shared display for physical meetings, finding that a shared display can reduce dominant behavior in in-person meetings \cite{dimicco2004influencing}. 

In the Meeting Mediator project, Kim et al. showed that real-time visualizations have the potential to increase interaction balance, decrease dominant behavior, and increase collaboration in physical meetings \cite{kim2008meeting}.
Later, the same system was shown to increase the effectiveness of distributed collaboration \cite{kim2012awareness}. 
While some of these works address distributed communication and real-time feedback, ours is the first to offer an open, accessible research tool for measuring group communication at scale and in real-world environments rather than laboratory experiments.

\section{Overview of the Breakout System}
The Breakout system (named after the concept of "breakout" groups in online learning) is comprised of three main technological components.
These components are described here as the \textbf{(1) data collection platform}, the \textbf{(2) analytics engine}, and the \textbf{(3) visualization platform}.
The data collection platform interfaces with modern video-conferencing tools to collect data on participants' actions while in a video chat. 
Breakout's analytics engine is a server-side component that is able to flexibly compute statistics and summary data of a group's interactions.
The visualization platform offers a modular visualization system that allows researchers to conceptualize and craft their own real-time interventions for online teams using data generated from the analytics engine.
The Breakout system is currently in beta release as a Google Hangout plugin, but has been designed for extensibility to other videoconference platforms.

\subsection{Data Collection Platform}
Prior works examining the group dynamics of teams has identified conversation dynamics as a core signal of group interaction and a major predictor of group performance \cite{woolley2010evidence}. 
A principal measurable signal of a team's conversation dynamics is its speaking history \cite{pentland2012new}. 
This information includes who is speaking, when they begin speaking, and for how long they speak.

For these reasons, the primary data collected by the platform is speaker history, excluding linguistic content of speech.
We use the API exposed by the video chat platform to determine when a user begins and ends a speaking period. The current platform examines the volume of participants' audio output to determine speaking times, disregarding short lower volume events as a rudimentary filter to increase data quality. 
Because each users' speaking volumes are recorded separately, this also allows us to identify simultaneous speaking events, or overlap, a signal that may reveal social relationships between participants, such as friendship \cite{dimicco2006using}.

The structure of distributed team communication may be more fluid than in-person groups, as participants can freely leave or join a conversation at will. 
Because of this dynamic, the platform also collects general participant event data from the video chat platform, such as when participants enter or leave a particular video chat. 
The data collected from this system is stored in a secure database, and is then subsequently made available as a real-time API, exposed through a REST interface as well as websockets.

\subsection{Analytics Engine}
The Analytics Engine is a small server-side application that allows analysis and aggregation of interaction and event data collected from the platform. 
It exposes a secure, real-time API, allowing authorized researchers and applications to access the computed data.
The Analytics Engine currently supports aggregate statistics such as (i) number of speaking events per person  (ii) response patterns between participants, made available as a directed graph where nodes are participants and edge weights correspond to the probability that the target node speaks after the source node (iii) frequency of turn-taking, a measure of engagement in team communication \cite{kim2012awareness}, and (iv) percentage of overlapped speaking time.

These statistics are computed at regular time intervals for all active videoconferences in the system. 
The computed data is then made available as a secure JSON API, and can be queried over HTTP or a websocket connection.

\subsection{Visualization Platform}
Breakout also includes a platform for creating modular visualizations based on collected data and the statistics generated by the analytics engine.
The Visualization Platform is implemented in Javascript, and utilizes D3.js, an industry-standard tool for data visualization \cite{bostock2011d3}. The platform allows for usage of basic aggregate measurements computed from the Analytics Platform, and enables the visualizations to update in real-time as the statistics are saved to the database. 

\begin{figure}[h]
    \begin{center}
    \includegraphics[width=0.7\textwidth]{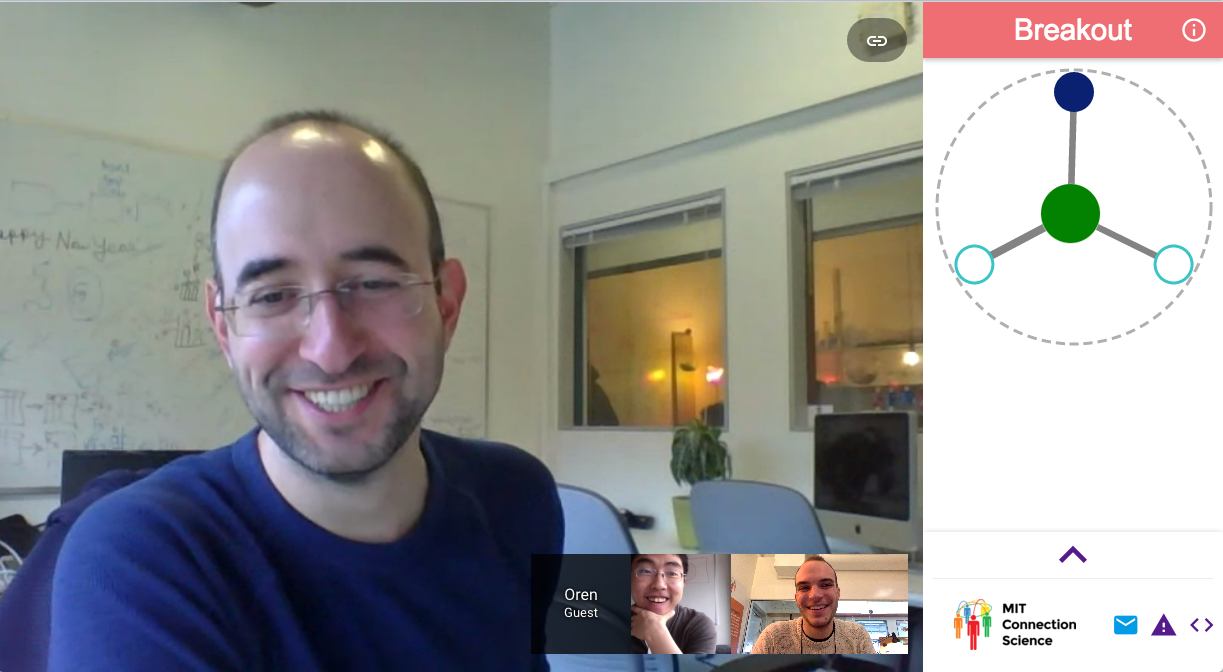}
    \caption{Screenshot of the visualization platform in a group videoconference on Google Hangouts}
    \label{fig:screenshot}
    \vspace{-3em}
    \end{center}
\end{figure}

\begin{figure}[t]
  \begin{center}
    \subfigure[]
    {
    \includegraphics[width=0.40\textwidth]{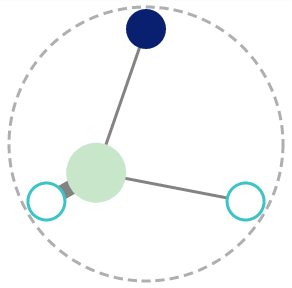}
    \label{fig:bad-group-viz}
    }
    \subfigure[]
    {
    \includegraphics[width=0.40\textwidth]{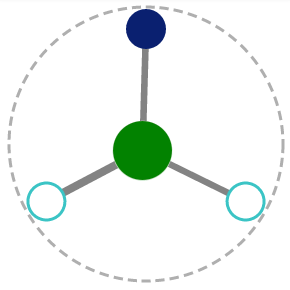}
    \label{fig:good-group-viz}
    }
  \end{center}
  \vspace{-2em}
  \caption{Examples of the custom Meeting Mediator (MM) visualization used in the current system. Figure \ref{fig:bad-group-viz} shows an example of the MM visualization in a group with one dominant member and low team engagement, while \ref{fig:good-group-viz} shows a group with high team engagement and no strongly dominant members.}
  \label{fig:visualizations}
\end{figure}

Figure \ref{fig:visualizations} shows an example visualization currently built-in to the system, a version of the Meeting Mediator (MM) visualization.
This visualization is based on prior group interaction research, and has been shown to improve group performance, trust, and engagement, while not increasing cognitive load \cite{kim2008meeting}.
In this visualization, each node on the outer edge of the circle represents a group member. 
As members of the group take more turns as a whole, the color of the green ball in the center becomes more intense. 
The ball moves closer to nodes (participants) who take more turns, offering a quick way of determining group members who are exhibiting dominant behavior.
Finally, the  thickness of the lines between participant nodes and the center ball corresponds with group members' levels of participation.

In the current implementation, visualizations such as these are displayed next to the video chat software, and update in real-time as the group members interact. 
Figure \ref{fig:screenshot} shows an example of a group meeting with the visualization platform enabled. 
In the current Google Hangout platform, visualizations appear on the right side of the group video chat.
This platform allows researchers to utilize real-time visualizations as an intervention strategy for online group communication studies. 
The modular nature of the system allows for quick development and deployment of prototype visualizations, which may quicken the pace of group dynamics and intervention research.

\section{Conclusion}
In this paper, we present Breakout, an open platform for online group interaction research based on the Google Hangout platform.
The platform allows researchers to easily collect data on group interaction, analyze the collected data, and deploy their own visualization-based interventions.
We expect that the usage of this tool in online learning contexts in particular will increase student engagement and offer an unprecedented opportunity to understand how group dynamics impact student learning outcomes.

\bibliographystyle{splncs03}
\bibliography{references}

\end{document}